\documentclass[12pt,a4paper,leqno]{article}
\usepackage{etoolbox} \usepackage{comment}
\usepackage{amsfonts}
\usepackage{multirow}
\usepackage{float}
\usepackage[makeroom]{cancel}
\usepackage[a4paper,includeheadfoot,margin=2.54cm]{geometry}
\usepackage{times}
\usepackage{booktabs}
\usepackage[amsthm,thmmarks]{ntheorem}
\usepackage{amsmath,amsfonts,amssymb,mathtools}
\usepackage{amsmath,xypic}
\usepackage{pdflscape}
\usepackage[export]{adjustbox} 
\usepackage[cal=boondoxo,calscaled=.96,scr=rsfs]{mathalpha} \usepackage{multicol}
\usepackage{hyphenat}
\usepackage[usenames,dvipsnames]{color}
\usepackage{rotating}
\usepackage{tikz-cd}
\usepackage{longtable}
\usepackage{caption}
\usepackage[most]{tcolorbox}

\usepackage{epigraph}
\usepackage{enumitem}
\setlist[enumerate]{listparindent=0.5in}
\usepackage{mathrsfs}
\newcommand{\be}{\begin{equation}}
\newcommand{\ee}{\end{equation}}
\newcommand{\bes}{\begin{equation*}}
\newcommand{\ees}{\end{equation*}}
\newcommand{\bea}{\begin{eqnarray}}
\newcommand{\eea}{\end{eqnarray}}
\newcommand{\beas}{\begin{eqnarray}}
\newcommand{\eeas}{\end{eqnarray}}
\newcommand{\ben}{\begin{note}}
\newcommand{\een}{\end{note}}
\newcommand{\bexl}{\vskip0.1em\noindent\hrulefill\vskip1em\begin{ExerciseList}}
\newcommand{\eexl}{\end{ExerciseList}\hrulefill}

\newcommand{\bthm}{\begin{theorem}}
\newcommand{\ethm}{\end{theorem}}
\newcommand{\bpro}{\begin{prop}}
\newcommand{\epro}{\end{prop}}
\newcommand{\bcor}{\begin{corollary}}
\newcommand{\ecor}{\end{corollary}}
\newcommand{\bcon}{\begin{conjecture}}
\newcommand{\econ}{\end{conjecture}}
\newcommand{\bp}{\begin{proof}}
\newcommand{\ep}{\end{proof}}
\newcommand{\blem}{\begin{lemma}}
\newcommand{\elem}{\end{lemma}}
\newcommand{\bn}{\begin{note}}
\newcommand{\en}{\end{note}}
\newcommand{\benum}{\begin{enumerate}}
\newcommand{\eenum}{\end{enumerate}}
\newcommand{\bed}{\begin{defn}}
\newcommand{\eed}{\end{defn}}
\newcommand{\brem}{\begin{remark}}
\newcommand{\erem}{\end{remark}}

\newcommand{\btik}{\begin{tikzpicture}\begin{axis}[scale=0.5,axis y line=center, axis x line=middle]}
\newcommand{\etik}{\end{axis}\end{tikzpicture}}

\let\into=\hookrightarrow

\newcommand{\upperRomannumeral}[1]{\uppercase\expandafter{\romannumeral#1}}

 \usepackage{graphicx}
\usepackage[pagewise]{lineno}
\usepackage{stackengine}
\usepackage{xr}
\usepackage{stmaryrd}
\usepackage[normalem]{ulem} \usepackage[cal=boondoxo,calscaled=.96,scr=rsfs]{mathalpha}
\DeclareMathAlphabet{\mathscrbf}{OMS}{mdugm}{b}{n}

\usepackage{titlesec}
\newtoggle{arxiv}
\toggletrue{arxiv}
\usepackage{natbib}
	\let\cite=\citep
\usepackage{fancyhdr}
	\usepackage[colorlinks,citecolor=blue]{hyperref}
\usepackage[capitalise,noabbrev]{cleveref}
\usepackage{fancyhdr}
\usepackage{cleveref}
\newtoggle{draft}
\togglefalse{draft}

\theoremstyle{theorem}
\newtheorem{theorem}[equation]{Theorem}      \newtheorem{lemma}[equation]{Lemma}          \newtheorem{corollary}[equation]{Corollary}  \newtheorem{proposition}[equation]{Proposition}

\newtheorem{conj}[equation]{Conjecture}

\newtheorem{defn}[equation]{Definition}

\theoremsymbol{\ensuremath{\bullet}}

\newtheorem{remark}[equation]{Remark}
\theoremsymbol{\ensuremath{\bullet}}

\numberwithin{equation}{section}

\let\oldproofname=\proofname
\renewcommand{\proofname}{{\bfseries\sffamily\textup{\oldproofname}}}

\titleformat{\subsection}[runin]{\normalfont\bfseries}{\thesubsection}{.5em}{}[{\ \ }]
\titlespacing{\subsection}{0pt}{1.5ex plus .1ex minus .2ex}{0pt}
\titleformat{\subsubsection}[runin]{\normalfont\bfseries}{\thesubsubsection}{.5em}{}[{\ \ }]
\titlespacing{\subsubsection}{0pt}{1.5ex plus .1ex minus .2ex}{0pt}

\newcommand{\nwss}{\numberwithin{equation}{subsection}}

\crefname{equation}{equation}{Equation}
\crefname{section}{§}{§§}
\crefname{subsection}{§}{§§}
\crefname{subsubsection}{§}{§§}

\usepackage{microtype}
\usepackage[margin=15pt,font=small,labelfont={bf,sf}]{caption}
\usepackage{sectsty}
\allsectionsfont{\sffamily}

\newcommand{\quotebox}[1]{
\begin{tcolorbox}[fontupper=\bfseries\itshape,fontlower=\normalfont,colback=white,boxrule=0.5pt]{#1}
\end{tcolorbox}
}

\newtcolorbox[auto counter,number within=subsection,
crefname={Table}{Table}]{boxedcontent}[2][]{colback=white,coltitle=black,colframe=white!25!brown,fonttitle=\bfseries,
	title=Table \thetcbcounter: #2,#1}

\let\into=\hookrightarrow
\let\isom=\simeq

\newcommand{\A}{\mathscr{A}}

\newcommand{\bF}{{\bar{F}}}
\newcommand{\bQ}{{\bar{\Q}}}

\newcommand{\C}{{\mathbb C}}

\newcommand{\F}{{\mathbb F}}

\newcommand{\N}{\mathscr{N}}

\newcommand{\Q}{{\mathbb Q}}
\newcommand{\R}{{\mathbb R}}

\renewcommand{\O}{{\mathscr O}}
\renewcommand{\P}{{\mathbb P}}
\renewcommand{\wp}{{\mathfrak p}}

\renewcommand{\bpro}{\begin{proposition}}
	\renewcommand{\epro}{\end{proposition}}
\renewcommand{\bcon}{\begin{conj}}
	\renewcommand{\econ}{\end{conj}}

\title{TuneGenie: Reasoning-based LLM agents for preferential music generation \\ \textcolor{red}{{\large\bf Preprint, under review}}}
\author{Amitesh Pandey, Jafarbek Arifdjanov, Ansh Tiwari}

\newcommand{\Address}{\bigskip\noindent{\footnotesize\textsc{{CMS. department, Caltech, 1200 E California Blvd., Pasadena
		CA 91126, USA.}}\par\nopagebreak 
\noindent\textit{Email:}	\texttt{amitesh@caltech.edu}}}

\begin{document}
	\maketitle
\epigraphwidth0.65\textwidth

\lhead{}

\iftoggle{draft}{\pagewiselinenumbers}{\relax}
\newcommand{\needsproof}{\textcolor{red}{{\large THIS NEEDS A PROOF\ }}}
\newcommand{\act}{\curvearrowright}
\newcommand{\lmp}{{\Pi\act\Ot}}
\newcommand{\lmpi}{{\lmp}_{\int}}
\newcommand{\lmpf}{\lmp_F}
\newcommand{\Om}{\O^{\times\mu}}
\newcommand{\Omf}{\O^{\times\mu}_{\bF}}
\renewcommand{\N}{\mathbb{N}}
\newcommand{\yoga}{Yoga}
\newcommand{\gl}[1]{{\rm GL}(#1)}
\newcommand{\bK}{\overline{K}}
\newcommand{\reptrip}{\rho:G_K\to\gl V}
\newcommand{\reptripp}[1]{\rho\circ\alpha:G_{\ifstrempty{#1}{K}{{#1}}}\to\gl V}
\newcommand{\benumlab}{\begin{enumerate}[label={{\bf(\arabic{*})}}]}
\newcommand{\benumlabresume}{\begin{enumerate}[resume,label={{\bf(\arabic{*})}}]}
\newcommand{\ord}{\mathop{\rm ord}\nolimits}	
\newcommand{\kcs}{K^\circledast}
\newcommand{\lcs}{L^\circledast}
\renewcommand{\A}{\mathbb{A}}
\newcommand{\bfq}{\bar{\mathbb{F}}_q}
\newcommand{\tripod}{\P^1-\{0,1728,\infty\}}

\newcommand{\vseq}[2]{{#1}_1,\ldots,{#1}_{#2}}
\newcommand{\anab}[4]{\left({#1},\{#3 \}\right)\anabelmap\left({#2},\{#4 \}\right)}

\newcommand{\gln}{{\rm GL}_n}
\newcommand{\glo}[1]{{\rm GL}_1(#1)}
\newcommand{\glt}[1]{{\rm GL_2}(#1)}

\newcommand{\iut}{\cite{mochizuki-iut1, mochizuki-iut2, mochizuki-iut3,mochizuki-iut4}}
\newcommand{\topics}{\cite{mochizuki-topics1,mochizuki-topics2,mochizuki-topics3}}

\newcommand{\linv}{\mathfrak{L}}
\newcommand{\bedef}{\begin{defn}}
\newcommand{\eedef}{\end{defn}}
\renewcommand{\act}[1][]{\overset{#1}{\curvearrowright}}
\newcommand{\bfx}{\overline{F(X)}}
\newcommand{\anabelmap}{\leftrightsquigarrow}
\newcommand{\ban}[1][G]{\mathscr{B}({#1})}
\newcommand{\pit}{\Pi^{temp}}
 
 \newcommand{\bL}{\overline{L}}
 \newcommand{\bkm}{\bK_M}
 \newcommand{\vbk}{v_{\bK}}
 \newcommand{\vbkm}{v_{\bkm}}
\newcommand{\ocs}{\O^\circledast}
\newcommand{\ot}{\O^\triangleright}
\newcommand{\ocsk}{\ocs_K}
\newcommand{\otk}{\ot_K}
\newcommand{\ok}{\O_K}
\newcommand{\oko}{\O_K^1}
\newcommand{\oks}{\ok^*}
\newcommand{\Qpb}{\overline{\Q}_p}
\newcommand{\Qpbh}{\widehat{\overline{\Q}}_p}
\newcommand{\tr}{\triangleright}
\newcommand{\ocpt}{\O_{\C_p}^\tr}
\newcommand{\ocpf}{\O_{\C_p}^\flat}
\newcommand{\sG}{\mathscr{G}}
\newcommand{\sX}{\mathscr{X}}
\newcommand{\sxfe}{\sX_{F,E}}
\newcommand{\sxfep}{\sX_{F,E'}}
\newcommand{\sxcpte}{\sX_{\cpt,E}}
\newcommand{\sxcptep}{\sX_{\cpt,E'}}
\newcommand{\loglt}{\log_{\sG}}
\newcommand{\fc}{\mathfrak{t}}
\newcommand{\ku}{K_u}
\newcommand{\kup}{\ku'}
\newcommand{\kt}{\tilde{K}}
\newcommand{\sGpf}{\sG(\O_K)^{pf}}
\newcommand{\hgm}{\widehat{\mathbb{G}}_m}
\newcommand{\bE}{\overline{E}}
\newcommand{\sY}{\mathscr{Y}}
\newcommand{\syfe}{\mathscr{Y}_{F,E}}
\newcommand{\syfep}{\mathscr{Y}_{F,E'}}
\newcommand{\syfqp}[1]{\mathscr{Y}_{\cptl{#1},\Q_p}}
\newcommand{\syfqpe}[1]{\mathscr{Y}_{{#1},E}}
\newcommand{\syfqpep}[1]{\mathscr{Y}_{{#1},E'}}
\newcommand{\fJ}{\mathfrak{J}}
\newcommand{\fM}{\mathfrak{M}}
\newcommand{\locvar}{local arithmetic-geometric anabelian variation of fundamental group of $X/E$ at $\wp$}
\newcommand{\fjxep}{\fJ(X,E,\wp)}
\newcommand{\fjxe}{\fJ(X,E)}
\newcommand{\fpc}[1]{\widehat{{\overline{\F_p(({#1}))}}}}
\newcommand{\cpt}{\C_p^\flat}
\newcommand{\cpti}{\C_{p_i}^\flat}
\newcommand{\cptl}[1]{\C_{p,{#1}}^\flat}
\newcommand{\fja}[1]{\fJ^{\rm arith}({#1})}
\newcommand{\ainfe}{A_{\inf,E}(\O_F)}
\renewcommand{\ainfe}{W_{\O_E}(\O_F)}
\newcommand{\gmh}{\widehat{\mathbb{G}}_m}
\newcommand{\sE}{\mathscr{E}}
\newcommand{\bpi}{B^{\varphi=\pi}}
\newcommand{\bpip}{B^{\varphi=p}}
\newcommand{\onto}{\twoheadrightarrow}

\newcommand{\cpmax}{{\C_p^{\rm max}}}
\newcommand{\xan}{X^{an}}
\newcommand{\yan}{Y^{an}}
\newcommand{\bPi}{\overline{\Pi}}
\newcommand{\bPit}{\bPi^{\rm{\scriptscriptstyle temp}}}
\newcommand{\Pit}{\Pi^{\rm{\scriptscriptstyle temp}}}
\renewcommand{\pit}[1]{\Pi^{\scriptscriptstyle temp}_{#1}}
\newcommand{\pitk}[2]{\Pi^{\scriptscriptstyle temp}_{#1;#2}}
\newcommand{\pio}[1]{\pi_1({#1})}
\newcommand{\fTeich}{\widetilde{\fJ(X/L)}}
\newcommand{\ssep}{\S\,} \newcommand{\vphi}{\varphi}
\newcommand{\sgt}{\widetilde{\sG}}
\newcommand{\sxqp}{\mathscr{X}_{\cpt,\Q_p}}
\newcommand{\syQp}{\mathscr{Y}_{\cpt,\Q_p}}

\setcounter{tocdepth}{2}

\newcommand{\mywork}[1]{\textcolor{red}{#1}}

\togglefalse{draft}
\newcommand{\FF}{\cite{fargues-fontaine}}
\iftoggle{draft}{\pagewiselinenumbers}{\relax}

\newcommand{\attportion}{Sections~\cref{se:number-field-case}, \cref{se:construct-att}, \cref{se:relation-to-iut}, \cref{se:self-similarity} and \cref{se:applications-elliptic}}

\newcommand{\inithtdata}{\cite[{\ssep\ref{III-ss:theta-data-fixing}, \ssep\ref{III-ss:theta-data-fixing2}}]{joshi-teich-rosetta}} 
\newcommand{\initassumptions}{Initial Theta Data \cite[{\ssep\ref{III-ss:elliptic-curve-assumptions}, \ssep\ref{III-ss:theta-data-fixing}, \ssep\ref{III-ss:theta-data-fixing2}}]{joshi-teich-rosetta}}
\newcommand{\assumptions}{\cref{sss:degrees}, \cref{sss:field-assump}}

\newcommand{\Pib}{\overline{\Pi}}
\newcommand{\four}{Sections~\cref{se:grothendieck-conj}, \cref{se:untilts-of-Pi}, and \cref{se:riemann-surfaces}}

\newcommand{\ENDDOCUMENT}{\phantomsection\addcontentsline{toc}{section}{References}\bibliography{references}\Address}

\begin{abstract}
Recently, Large language models (LLMs) have shown great promise across a diversity of tasks ranging from generating images to reasoning spatially. Considering their remarkable (and growing) textual reasoning capabilities, we investigate LLMs' potency in conducting analyses of an individual's preferences in music (based on playlist metadata, personal write-ups, etc.) and producing effective prompts (based on these analyses) to be passed to Suno AI (a generative AI tool for music production). Our proposition of a novel LLM-based textual representation $\to$ music model (which we call \emph{TuneGenie}) and the various methods we develop to evaluate \& benchmark similar models add to the increasing (and increasingly controversial) corpus of research on the use of AI in generating art. 
\end{abstract}

\numberwithin{equation}{subsection}
\newcommand{\omu}{\O_{\bQ_p}^{\mu}}
\newcommand{\lmod}{L_{\rm mod}}
\newcommand{\ttheta}{{{\widetilde\Theta}}}
\newcommand{\tthetaj}[1]{\ttheta_{\scriptscriptstyle{Joshi},#1}}
\newcommand{\tthetam}[1]{\ttheta_{\scriptscriptstyle{Mochizuki},#1}}

\newcommand{\moccor}{\cite[Corollary 3.12]{mochizuki-iut3}}
\newcommand{\thetam}{{\ttheta}_{\scriptscriptstyle{Mochizuki}}}
\newcommand{\thetamp}{{\ttheta}_{\scriptscriptstyle{Mochizuki},p}}
\newcommand{\thetampi}{{\ttheta}_{\scriptscriptstyle{Mochizuki},p}^{\bsI}}
\newcommand{\thetaminfty}{{\ttheta}_{\scriptscriptstyle{Mochizuki},\infty}^{\bsI}}
\newcommand{\thetaj}{{{\ttheta}_{\scriptscriptstyle{Joshi}}}}
\newcommand{\thetajp}{{{\ttheta}_{\scriptscriptstyle{Joshi},p}}}
\newcommand{\sM}{\mathscr{M}}
\newcommand{\sMb}{\overline{\mathscr{M}}}
\newcommand{\pib}{\overline{\Pi}}
\newcommand{\bN}{\mathbb{N}}
\newcommand{\sD}{\mathscr{D}}
\newcommand{\sF}{\mathscr{F}}
\newcommand{\sL}{\mathscr{L}}

\newcommand{\bdrp}{B_{dR}^+}
\newcommand{\bdr}{B_{dR}}

\newcommand{\iutthr}{\cite{mochizuki-iut1,mochizuki-iut2, mochizuki-iut3}}

\newcommand{\thetajpp}{{\widehat{\Theta}}_{\scriptscriptstyle{Joshi},p}}
\newcommand{\thetaja}{{\widehat{\Theta}}_{\scriptscriptstyle{Joshi}}}
\newcommand{\thetajpph}{{\widehat{\widehat{\Theta}}}_{\scriptscriptstyle{Joshi},p}}
\newcommand{\thetajppa}{{\widehat{\widehat{\Theta}}}_{\scriptscriptstyle{Joshi}}}

\newcommand{\ells}{{\ell^*}}
\newcommand{\sP}{\mathscr{P}}
\newcommand{\spc}{\sP^{Teich}}
\newcommand{\pitop}[1]{\pi_1^{top}(#1)}
\newcommand{\sppi}{\sP(\Pi\hookleftarrow\pib)}
\newcommand{\sppim}{\sP_{\scriptscriptstyle{Mochizuki}}(\Pi\hookleftarrow\pib)}
\newcommand{\sppij}{\sP_{\scriptscriptstyle{Joshi}}(\Pi\hookleftarrow\pib)}

\newcommand{\bcris}{B_{cris}}

\newcommand{\hol}[3]{\mathfrak{hol}_{#1}(#3)_{#2}}

\newcommand{\xdia}{X^\lozenge}
\newcommand{\holdia}{{\mathfrak{Hol}(X/E)}}
\newcommand{\spd}{{\rm Spd}}
\newcommand{\perf}{{\rm Perf}}
\newcommand{\perfld}{{\rm PerfFld}}
\newcommand{\spa}{{\rm Spa}}
\newcommand{\bnddsub}{{\tiny\circ}}
\newcommand{\xad}{X^{ad}}
\newcommand{\sppimold}{\sP_{\scriptscriptstyle{Mochizuki}}(\Pi\hookleftarrow\pib)}
\renewcommand{\sppim}{\sP_{\scriptscriptstyle{Mochizuki}}'(\Pi\hookleftarrow\pib)}

\newcommand{\chxfe}{{\rm Ch^1(\sxfe)}}
\newcommand{\divxfe}{{\rm Div}(\sxfe)}

\newcommand{\qprojqp}[1]{\mathcal{QProj}_{#1}}
\newcommand{\projqp}[1]{\mathcal{Proj}_{#1}}

\newcommand{\tsigma}{\tilde{\sigma}}
\newcommand{\ismg}{{\mathbf{Ism}(G)}}
\renewcommand{\tripod}{\P^1-\{0,1,\infty\}}
\newcommand{\okbt}{\O_{\bE;K}^\triangleright}
\newcommand{\Ob}{\overline{\O}} 

\renewcommand{\deg}{{\rm deg}}
\newcommand{\udeg}[1]{\underline{{\rm deg}}(#1)}
\newcommand{\divp}{Div_{+}}
\newcommand{\divt}{\widetilde{{\rm Div_{+}}}}
\newcommand{\pdivt}{\widetilde{{\rm PDiv}}}
\newcommand{\fjxlp}{\fJ(X,L_\wp)}

\newcommand{\bbvl}{\mathbb{V}_L}
\newcommand{\bbvlp}{\mathbb{V}_{L'}}
\newcommand{\bbvlmod}{\mathbb{V}_{\lmod}}
\newcommand{\bbvlmodgood}{\mathbb{V}_{\lmod}^{good}}
\newcommand{\bbvlgood}{\mathbb{V}_{L}^{good}}
\newcommand{\bbvlpgood}{\mathbb{V}_{L'}^{good}}
\newcommand{\bbvlmodoss}{\mathbb{V}_{\lmod}^{odd,ss}}
\newcommand{\bbvloss}{\mathbb{V}_{L}^{odd,ss}}
\newcommand{\bbvlposs}{\mathbb{V}_{L'}^{odd,ss}}
\newcommand{\ubblv}{\underline{\mathbb{V}}}
\newcommand{\ubblvgood}{\underline{\mathbb{V}}^{good}}
\newcommand{\ubblvoss}{\underline{\mathbb{V}}^{odd,ss}}
\newcommand{\ubblvossp}{\underline{\mathbb{V}}^{odd,ss}_p}

\newcommand{\ul}[1]{\underline{#1}}

\newcommand{\breveB}{{\breve{B}}}
\newcommand{\breveBb}{{\breve{\mathbf{B}}}}
\newcommand{\bigB}{\bigotimes_{E'\anabelmap E} \left(\oplus_{j=1}^\ells B_{E'} \right)}
\newcommand{\tSigma}{\widetilde{\Sigma}}
\newcommand{\thetajh}{{\widehat{\Theta}}_{\scriptscriptstyle{Joshi}}}

\newcommand{\sI}{\mathcal{I}}
\newcommand{\utheta}{U_\Theta}

\newcommand{\logsh}[1]{\mathscr{I}({#1})}
\newcommand{\logshq}[1]{\mathscr{I}^{\Q_p}({#1})}
\newcommand{\logshe}{\logsh{E}}
\newcommand{\logsheq}{\logshq{E}}
\newcommand{\logshep}{\logsh{E'}}
\newcommand{\logshepq}{\logshq{E'}}

\newcommand{\bS}{\mathbb{S}}

\newcommand{\sIp}{\sI_p}
\newcommand{\sIpq}{{\sIp}^{\Q}}
\newcommand{\sIm}{\sI_{\scriptscriptstyle{Mochizuki}}}
\newcommand{\sImq}{\sI_{\scriptscriptstyle{Mochizuki}}^{\Q}}
\newcommand{\mulog}[2]{\mu_{#1}^{\log}({#2})}
\newcommand{\EB}{{}^EB} 

\newcommand{\vnon}{\V^{non}}
\newcommand{\varc}{\V^{arc}}

\newcommand{\dmod}{d_{mod}}
\newcommand{\dmods}{d^*_{mod}}
\newcommand{\emod}{e_{mod}}
\newcommand{\emods}{e^*_{mod}}
\newcommand{\ltpd}{L_{tpd}}
\newcommand{\ff}{\mathfrak{f}}
\newcommand{\fd}{\mathfrak{d}}
\newcommand{\fq}{\mathfrak{q}}
\newcommand{\fQ}{\mathfrak{Q}}
\newcommand{\fT}{\mathfrak{Tate}}
\newcommand{\fl}{\mathfrak{l}}
\newcommand{\dtpd}{\fd_{\ltpd}}
\newcommand{\ftpd}{\ff_{\ltpd}}
\newcommand{\etap}{\eta_{prm}}
\newcommand{\V}{\mathbb{V}}
\newcommand{\vq}{\V_{\Q}}
\newcommand{\vnl}{\V_L^{non}}
\newcommand{\vlnon}{\vnl}
\newcommand{\vlarch}{\mathbb{V}_L^{arc}}
\newcommand{\vlp}{\V_{L'}}
\newcommand{\vlpp}{\V_{L',p}}
\newcommand{\vnlp}{\V_{L'}^{non}}
\newcommand{\vlnonp}{\vnlp}
\newcommand{\vlarchp}{\mathbb{V}_L^{'arc}}
\newcommand{\fA}{\mathfrak{A}}

\newcommand{\bsI}{\boldsymbol{\mathscrbf{I}}}
\newcommand{\thetajti}{{\thetaj}^{\tilde{\mathscrbf{I}}}}
\newcommand{\thetaji}{{\thetaj}^{\mathscrbf{I}}}
\newcommand{\thetamti}{{\thetam}^{\tilde{\mathscrbf{I}}}}
\newcommand{\thetami}{{\thetam}^{\mathscrbf{I}}}
\newcommand{\flog}{\mathfrak{log}}

\newcommand{\lmodv}{L_{mod,v}}
\newcommand{\vmdst}{\V_M^{dst}}
\newcommand{\vqdst}{\V_{\Q}^{dst}}
\newcommand{\vmnon}{\vnon_M}
\newcommand{\supp}{{\rm Supp}}

\newpage
{\sffamily
\hypersetup{linkcolor=red}
	\tableofcontents
}

\newpage

\definecolor{darkmidnightblue}{rgb}{0.0, 0.2, 0.4}
\definecolor{carmine}{rgb}{0.59, 0.0, 0.09}
\newtcbox{\mybox}[1][red]{on line,
	arc=0pt,outer arc=0pt,colback=white,colframe=darkmidnightblue,
	boxsep=0pt,left=0pt,right=0pt,top=2pt,bottom=2pt,
	boxrule=0pt,leftrule=1pt, rightrule=1pt,bottomrule=1pt,toprule=1pt}
\newcommand{\tmb}[1]{\mybox{#1}}
\newcommand{\present}{present series of papers (\cite{joshi-anabelomorphy,joshi-untilts,joshi-teich-estimates,joshi-teich-def,joshi-teich-rosetta}  and \cite{joshi-anabelomorphy,joshi-formal-groups})}

\newcommand{\abc}{$abc$-conjecture}
\epigraph{Damn, you wimps$\dots$\\
I thought you wanted to \emph{change the world}!?}{Unknown}
\section{Introduction}\label{se:intro}\nwss

\newcommand{\vl}{\V_L}
\newcommand{\lvbh}{\widehat{\overline{L}}_v}
\newcommand{\lvbht}{{\widehat{\overline{L}}_v}^\flat}
\newcommand{\lvbhmax}{\widehat{\overline{L}}_v^{max}}
\newcommand{\lvbhtmax}{{\widehat{\overline{L}}_v}^{max\flat}}
\newcommand{\lvbhreal}{\widehat{\overline{L}}_v^{\R}}
\newcommand{\lvbhtreal}{{\widehat{\overline{L}}_v}^{\R\flat}}

\newcommand{\arith}[1]{\mathfrak{arith}(#1)}
\newcommand{\arithl}{\arith{L}}
\newcommand{\adel}[1]{\mathfrak{adel}(#1)}
\newcommand{\adell}{\adel{L}}
\newcommand{\by}{{\bf y}}

\newcommand{\syflv}{\sY_{\lvbht,L_v}}
\newcommand{\syflvmax}{\sY_{\lvbhtmax,L_v}}
\newcommand{\sxflv}{\sX_{\lvbht,L_v}}
\newcommand{\sxflvmax}{\sX_{\lvbhtmax,L_v}}
\newcommand{\bsY}{\mathscrbf{Y}}
\newcommand{\bsX}{\mathscrbf{X}}
\newcommand{\yadl}{\bsY_L}
\newcommand{\yadlp}{\bsY_{L'}}
\newcommand{\yadlmax}{\bsY_L^{max}}
\newcommand{\yadq}{\bsY_{\Q}}
\newcommand{\yadqmax}{\bsY_{\Q}^{max}}
\newcommand{\xadl}{\bsX_L}
\newcommand{\xadq}{\bsX_{\Q}}
\newcommand{\xadlmax}{\bsX_L^{max}}
\newcommand{\xadqmax}{\bsX_{\Q}^{max}}
\newcommand{\yadlpoint}{\{(L_\wp\into K_{\wp}, K_\wp\isom \cpt)\}_{\wp\in\vlnon}}
\newcommand{\xadlpoint}{\{(L_\wp\into K_{\wp}, K_\wp\isom \cpt)\}_{\wp\in\vlnon}}

\newcommand{\bv}{\bar{v}}
\newcommand{\sGal}[1]{\mathbf{G}_{#1}}

\newcommand{\fet}{\mathcal{F\hat{e}t}}
\newcommand{\sA}{\mathcal{A}}

\newcommand{\logp}{\log^+}
\newcommand{\sFrob}{\mathcal{Frob}}
\newcommand{\pow}[2]{#1\llbracket#2\rrbracket}

\newcommand{\ainf}{A_{\textrm{inf}}}

\newcommand{\xs}{X^S}
\newcommand{\sZ}{\mathscr{Z}}
\newcommand{\bvphi}{\boldsymbol{\varphi}}
\subsection{Motif}
The field of Generative AI (GenAI) has bustled with recent developments in deep generative models for music and sound \cite{dhariwal2020jukeboxgenerativemodelmusic}, and even more importantly (and popularly) with large language models (LLMs) \cite{cao2023comprehensivesurveyaigeneratedcontent}. Much research has been conducted in attempts to utilize LLMs' remarkable abilities (by inducing reasoning through chains-of-thought in spatial contexts \cite{wei2023chainofthoughtpromptingelicitsreasoning}\cite{yamada2024l3golanguageagentschainof3dthoughts}, in image-splicing \cite{nath2025chatgptperformimagesplicing}, and even in crafty representations of financial assets \cite{sarkar}). It stands to reason that, with these unexpected applications arising at the intersection of disparate domains, it is worth exploring (and developing) cohesion between representations of an individual's taste in music and generative music models (like Udio or Suno\footnote{Our principal tool of use: \texttt{https://www.suno.com/home}}) through the \textcolor{red}{\emph{fabric of LLMs}}. 
\subsection{Recent \& Related Work}
\textbf{LLMs + Existing Deep Generative Models for Music}
\vspace{0.2cm}
\\ In an early study \cite{wei2025contextaitunescontextawareaigenerated}, Wei \emph{et al.} already explore developing \& integrating LLM-generated prompts to Suno in different environmental contexts tailored for stress relief benefits. The exact problem they pose (to the community) and investigate is: \quotebox{Does AI-generated music tailored to the user’s environment and stress level outperform pre-recorded relaxing music in effectiveness?}\vspace{0.15cm}
More generally, Ding \emph{et al.} explore composing lyrics and melodies using symbolic representations of songs as opposed to wavetransforms per usual. These representations themselves are entirely empirical in nature, making use of encodings with the $\log$-transform on a time series relative to its starting point. They employ LLMs primarily in the continuation of songs using these learned representations \cite{ding2025songcomposerlargelanguagemodel}.\\\\
\textbf{LLMs in detecting GenAI music}\vspace{0.2cm}
\\There is controversy and a general sense of alarm surrounding GenAI + Art (/music). To highlight and attempt to resolve this issue, in \cite{rahman2025sonicssyntheticidentifying}, methods are developed that reportedly excel at detecting ``counterfeit" music, generated through the means of Suno, and could presumably work for even more foundational models (like \cite{gong2025acestepstepmusicgeneration}).\newpage
\noindent{\textbf{Q\&A, Query, and Audio Datasets}}\vspace{0.2cm}
\\Tangential to our work and having the potential to cast a relatively unexplored space when combined with our work is the work of Deng\emph{ et al.} in \emph{MusiLingo}, specifically, the \verb|MI| dataset, and the question of if it can be utilised in improving generated prompts by improving the underlying LLM's reasoning capacity. \cite{deng2024musilingobridgingmusictext}. \\\\
There are a number of large, meticulously curated, and high-quality datasets used typically in ML + audio settings. These include the pure melody-based dataset \cite{raffel2016learning} (which can be worked with by conversion to MIDI), the recent LP-MusicCaps-MSD dataset primarily for song captioning \cite{doh2023lp}, and the single-note-based NPSynth dataset \cite{engel2017neural}. \\\\
\textbf{Abstractions and Representations}\vspace{0.2cm}
\\In relation to our work on formalising individuals' music preferences' representations, Anazbaghan \emph{et al.} probe LLMs to work purely on rudimentary textual representations with relative success \cite{anbazhagan2025probingaudiogenerationcapabilitiestextbased}. This serves as a natural baseline for our work to improve upon. \\\\
In \cite{sarkar}, \cite{hu2018deepstockrepresentationlearning}, representations of financial assets covered are particularly inspiring. There is a somewhat unintuitive parallel to be drawn here:
\begin{itemize}
    \item Candle stick CAE-image $\to$ 1D Audiowaveform's Spectrogram CAE-based image
    \item News Reports/Investor sentiments  $\to$ Music enthusiasts' review
    \item Company legal filings/press releases $\to$ individual's preferences
\end{itemize}
Also noteworthy is the prior discussion on the symbolic encoding method in \cite{ding2025songcomposerlargelanguagemodel}.
\subsection{Novelty Introduced}
TuneGenie introduces several key novel methods:
\begin{enumerate}
    \item \textbf{Novel Representation Method: }Motivated by the growing body of work on developing representations of financial assets using both ``more traditional" ML techniques \cite{hu2018deepstockrepresentationlearning} and LLMs \cite{sarkar}, we introduce a method to create an individual's representation in the context of the songs they prefer to listen to.
    \item \textbf{Cohesive Integration Method: }The traversal of extracting relevant data from playlists, etc., to LLM-generated prompts, and finally through the \verb|suno-api|, to music is non-trivial and unexplored. Admittedly, the work in \cite{wei2025contextaitunescontextawareaigenerated} comes close.
    \item \textbf{Robust Testing through Similarity Measures: }We test various similarity measures in pitting the generated music against the individual's preferred music to gauge the performance of our model. Song similarity has been explored in counterfeit-music detection, as in \cite{rahman2025sonicssyntheticidentifying}, but not in this ``noisy \& open-ended" context.
\end{enumerate}\newpage
\section{Methodology}
TuneGenie is implemented to replicate the seamlessness, low computational load, and ease in local execution of an end-to-end product. The computational and storage load is offset by reliance on Telegram's and Google's cloud services, which are available freely to all users. In a business setting, however, this would have to be endured by the producers. We incorporate an additional verifying agent into the prompt generator to add more accountability to our model.
\subsection{Model Architecture}
\begin{figure}[ht]
    \centering
    \includegraphics[width=\linewidth]{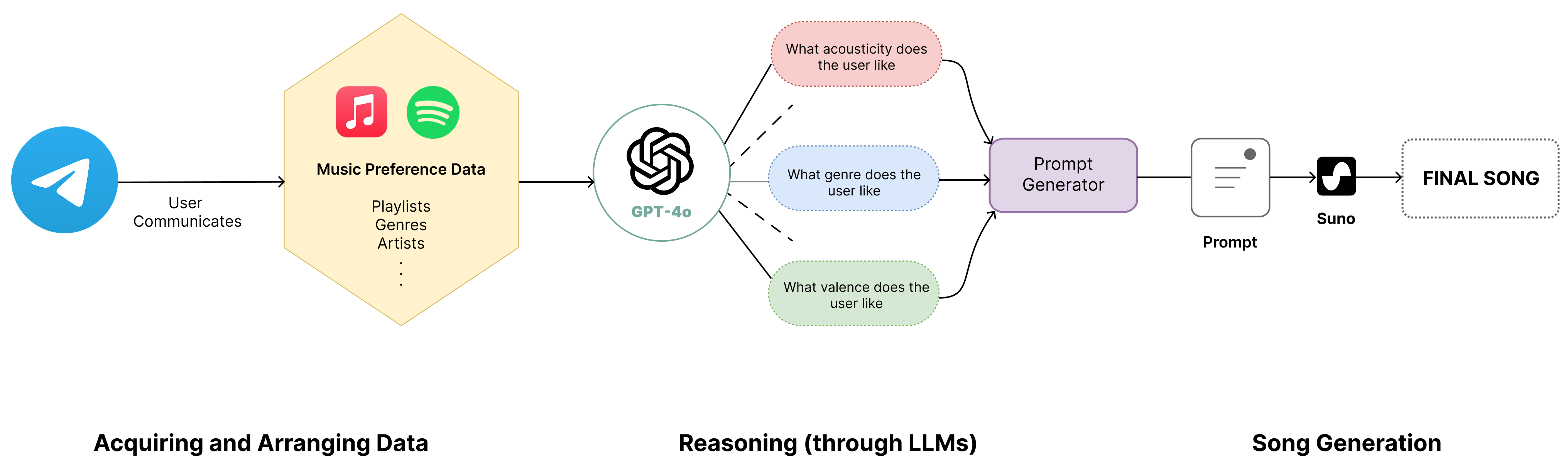}
    \caption{Model Framework}
    \label{fig:enter-label}
\end{figure}
\noindent We now provide a brief description of each transition:
\begin{itemize}
    \item Telegram $\to$ Preference Data: To alleviate the computational resources expenses, at the moment, Telegram is used to facilitate going from a user's playlist, liked songs, followed artists, etc., to our data format.
    \item Preferred Data $\to^{*}$ LLM: The collated preferred data provided by the user is first converted into a representation of the user, which is supposed to improve the ease with which the LLM can handle this data.
    \item LLM $\to$ Forced Reasoning: This is a \emph{less rigorous} way of self-ask prompting \cite{press2023measuringnarrowingcompositionalitygap}, and is supposed to let the LLM internalise relevant inferences from representation level data. We pose a variety of questions, available in the appendix.
    \item Reasoned LLM State $\to$ Prompt Generation: Here, a \emph{verifying agent} is introduced, and detailed further in Figure 2. To contribute robustness to the prompts, we use the LLM's reasoning abilities to verify the correctness of these prompts given the user's preferences.
    \item Prompt $\to$ Final Song: Unfortunately, Suno AI \emph{does not offer} an API. To counter this issue, we used Selenium + ChromeDriver to automate the process of inputting the prompts. This is only possible in an academic setting. In the real world, a business license with Suno would likely need to be purchased.
\end{itemize}
\subsection{Representation Paradigm}
The dialogue surrounding a particular song or an artist carries crucial information about the sentiment that surrounds a set of songs. On the other hand, an individual's sentiment towards a set of songs or artists reflects some inherent characteristic. These are two sides of the same coin: matching music to man.  \\\\
The critical problem is that there is no structure to the various discourses on album releases on social media or press coverage of musical artists. There is also no structure to the primary sources that the industry collects for musical data—for instance, YouTube comments on an album. We create embeddings, which are multi-dimensional vectors of representation.  For each song and each user, there is a representative vector. For each song $s$, user $u$, the interaction:
\begin{equation*}
\mathbf r_{s, u}=\overbrace{\Delta \boldsymbol{v}_{s} \cdot \boldsymbol{x}_{u}}^{\text {song contribution }}+\overbrace{\boldsymbol{v}_{s} \cdot \Delta \boldsymbol{x}_{u}}^{\text {user contribution }}+\varepsilon_{s, u}
\end{equation*}
Then, for user $u$ with $\deg(u)=k$, for example, the final pass $\mathbf{p}$ to the LLM is simply:
\begin{equation*}
    \mathbf{p}_u = \frac{1}{k}\sum_{i=0}^{k}\mathbf{r}_{i,u}
\end{equation*}
This is essentially a graphical approach to modelling the problem.
\begin{center}
    \begin{tikzpicture}[
    scale=0.6,
    node/.style={circle, draw, minimum size=0.8cm, fill=white},
    bridge/.style={circle, draw, minimum size=0.8cm, fill=red!20},
    edge/.style={-}
]

\node[node] (A1) at (0, 2) {$s_1$};
\node[node] (A2) at (1.5, 3) {$s_2$};
\node[node] (A3) at (1.5, 1) {$s_3$};
\node[node] (A4) at (3, 2) {$s_4$};

\node[node] (B1) at (8, 3) {$u_1$};
\node[node] (B2) at (9.5, 4) {$u_2$};
\node[node] (B3) at (9.5, 2) {$u_3$};
\node[node] (B4) at (11, 3) {$u_4$};
\node[node] (B5) at (8, 1) {$u_5$};

\node[bridge] (X) at (5.5, 0) {$u$};

\draw[edge] (A1) -- (A2);
\draw[edge] (A1) -- (A3);
\draw[edge] (A2) -- (A4);
\draw[edge] (A3) -- (A4);
\draw[edge] (A2) -- (A3);

\draw[edge] (B1) -- (B2);
\draw[edge] (B1) -- (B3);
\draw[edge] (B2) -- (B4);
\draw[edge] (B3) -- (B4);
\draw[edge] (B1) -- (B5);
\draw[edge] (B3) -- (B5);

\draw[edge, thick, blue] (X) -- (A1);
\draw[edge, thick, blue] (X) -- (A3);
\draw[edge, thick, blue] (X) -- (A4);

\draw[edge, thick, blue] (X) -- (B1);
\draw[edge, thick, blue] (X) -- (B3);
\draw[edge, thick, blue] (X) -- (B5);

\node[font=\small] at (1.5, 4.5) {Songs};
\node[font=\small] at (9.5, 5.5) {Listeners/Users};
\node[font=\small] at (5.5, -1.3) {Primary User};

\end{tikzpicture}
\end{center}
Intuitively, what we are doing is taking the weight of each edge in the graph for the current operating user (this can be a user-user or user-song interaction) and averaging it out. Over some samples of components (for a set of users, a set of songs of one genre, etc.), this creates a unique identity for a user. Finally, this is used to improve the LLM's prompt-generation process.
\begin{table}[h]
\centering
\caption{Data Sources}
\label{tab:datasources}
\begin{tabular}{@{}clp{10cm}@{}}
\toprule
\textbf{ID} & \textbf{Data Type} & \textbf{Data Source} \\
\midrule
1 & Social Media & YouTube Comments, Instagram Posts, Reddit Communities \\
2 & Streaming Platforms & YouTube Music, Spotify Plays, Apple Music Follows \\
3 & Personal Preference & Self-described, sentiment encoded \\
\bottomrule
\end{tabular}
\end{table}

\subsection{Seamless Scoring with Unsupervised Methods}
The idea of similarity is not new to audio signals, as it is of relevance in recommender systems, counterfeit detection and visualisation in user interfaces (audio visualisers). Recall how the rudimentary measures of similarity (in unsupervised learning) work. For example, in $k$-means \cite{macqueen1967some}, a method to classify data based on similarity in $k$ clusters, the objective:
\begin{equation*}
J = \sum_{i=1}^{k} \sum_{x \in C_i} \|x - \mu_i\|^2 \text{ is minimized}
\end{equation*}
where $C_i$ is the $i$-th cluster and $\mu_i$ is the centroid of cluster $C_i$.
The algorithm iteratively updates cluster centroids and reassigns points. The centroid of each cluster is computed as:
\begin{equation*}
\mu_i = \frac{1}{|C_i|} \sum_{x \in C_i} x
\end{equation*}
where $|C_i|$ denotes the number of points in cluster $C_i$.

Each data point $x_j$ is assigned to the cluster with the nearest centroid:
\begin{equation*}
c_j = \arg\min_{i} \|x_j - \mu_i\|^2
\end{equation*}
using Euclidean distance $d(x, \mu_i) = \sqrt{\sum_{d=1}^{p} (x_d - \mu_{i,d})^2}$.
The algorithm converges when centroid changes fall below a threshold:
\begin{equation*}
\sum_{i=1}^{k} \|\mu_i^{(t+1)} - \mu_i^{(t)}\| < \epsilon
\end{equation*}
Notice that we can use $k$-means directly with our representations of songs, as we work with vectors. In particular, we can create a representation of the generated song (using available markers) and measure its distance to the preferred cluster. As a rudimentary example:
\begin{figure}[h]
    \centering
    \includegraphics[width=0.7\linewidth]{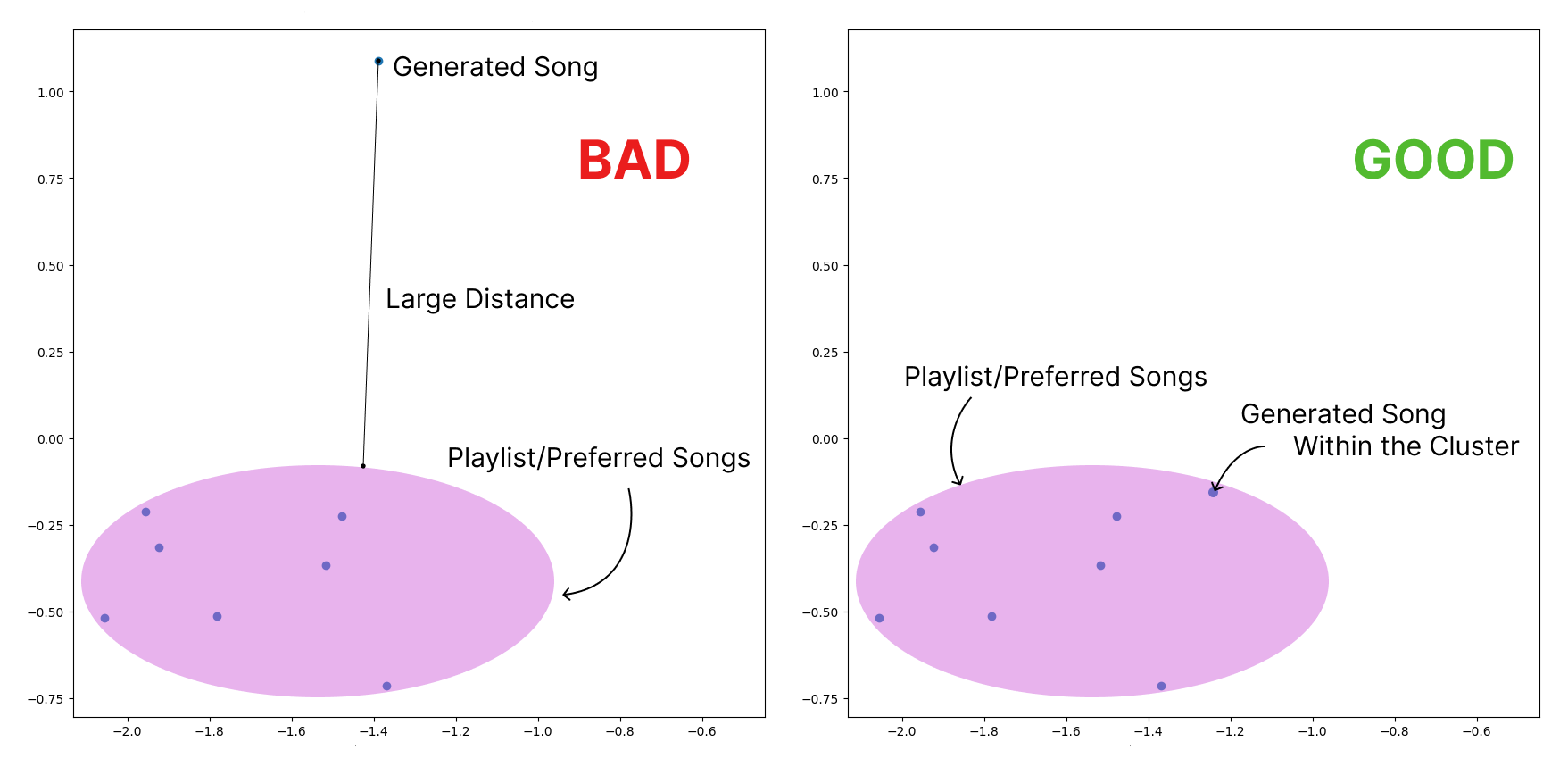}
    \caption{2-axis song clustering}
    \label{fig:enter-label}
\end{figure}
\newpage
\subsection{Experiment with Volunteers}
By far, the strongest way to test our model is by randomly recruiting volunteers (users) and presenting them with a questionnaire and music streaming profiles to create their user representations. Then, presenting them with some sample of the generated music. 
\section{Results}
\subsection{Examples}
Have a listen to \href{https://suno.com/song/ad284fdf-23b7-4c32-b345-532e5491626e?sh=kFbO1nwHNFTQaT1c}{this}, \href{https://suno.com/song/ec475db2-5a81-409f-b391-e641088013dd?sh=Bb4OcMMJIbP3ALSB}{this}, and also \href{https://suno.com/song/82702d5b-bf9f-4235-b5d8-bf1653811535?sh=KxR4Fle8dsi4G5Ri}{this}. These were based on the playlist attached in the appendix and run on the 2nd author's representation. These are impressive results.
\subsection{Similarity Results}
We plotted the generated songs' dimension \& information-``reduced" representations to measure similarity. Here are some plots of different cases. Playlists used are in the appendix.
\begin{figure}[h]
    \centering
    \includegraphics[width=\linewidth]{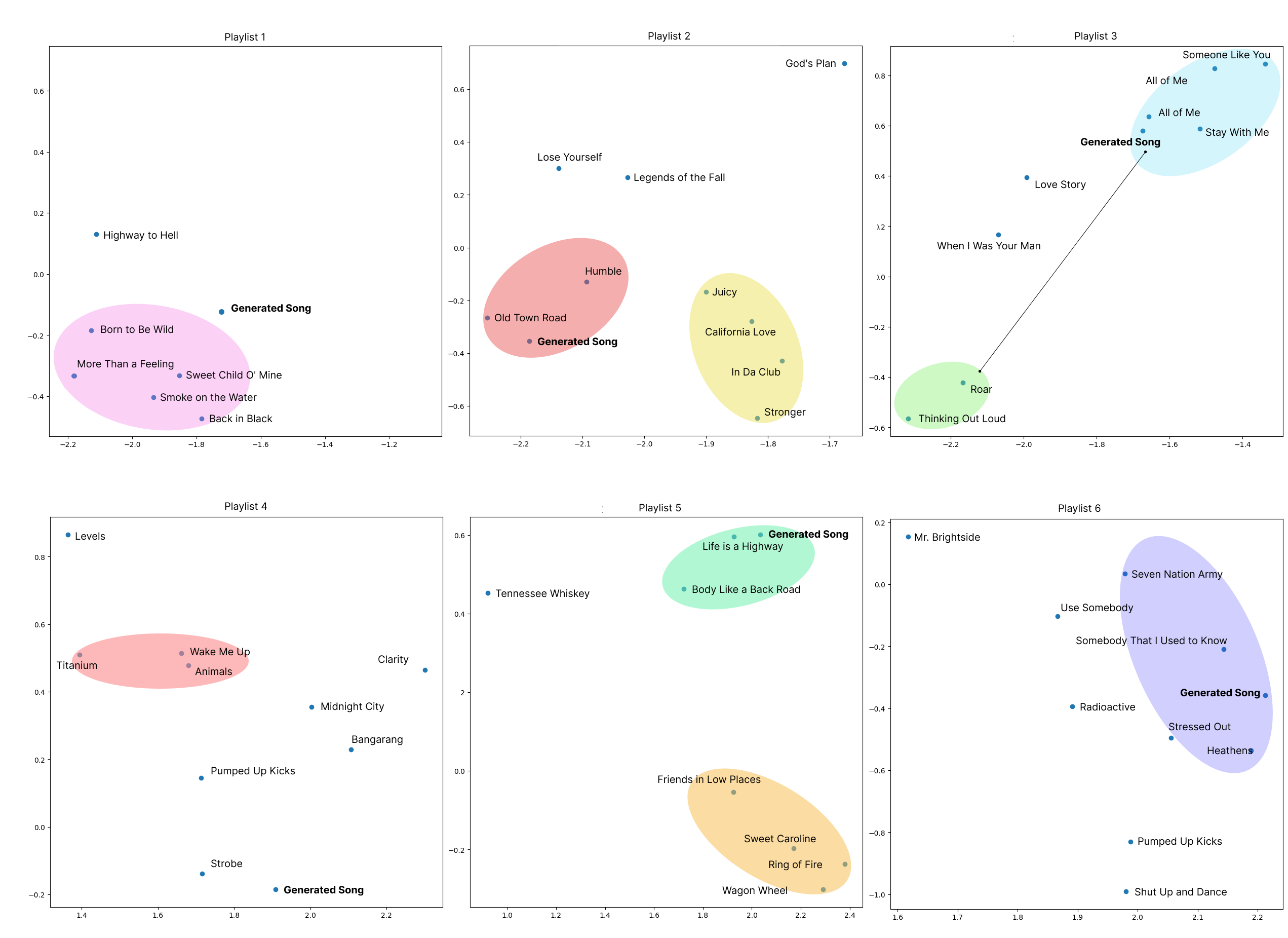}
    \caption{SVD Clustering Results}
    \label{fig:enter-label}
\end{figure}\\
As can be seen, overall, the generated song manages to place within the overarching cluster.
\newpage
\subsection{Volunteer Experiment Results}
\textcolor{red}{Work in Progress}
\section{Conclusion}
We introduced TuneGenie, an open-source low-cost model for preferential music generation. This body of work manages to successfully illustrate  the potency of our LLM-backed architecture, which integrates a reasoning setup, a verification module, and an audio generative model together. This highlights a major step in the direction of realizing a robust song generation foundation model that can reason with the user, a parallel not to be found in even other domains, such as image generation. We are confident that the gap between open-source and closed-source models in music generation can
be progressively narrowed leveraging our work.
\section{Limitations}
The first limitation is the reliance on Suno AI, which is a closed source, and proprietary audio generation software. Outside the academic sense, TuneGenie is not applicable owing to its \emph{breach of Suno AI's terms \& conditions.} However, even with a business license of the sort that allows for this work, the lack of a Suno API necessitates a the need for a webscript using Selenium which is not formally applicable and poses high computational loads. A user interface that interacts directly with the user as opposed to an interlay backed by Telegram would dispel some of these computational costs, however it would induce the need for a server compute separate from Google's or Telegram's. The representation used can be modified to be more effective for SVD/PCA/$k$-means methods to make the similarity measurement more accurate.
\section{Credits}
This project was conducted as part of Caltech's CS 159: Advanced Topics in Machine Learning (LLMs for Reasoning) course in Spring 2025. We'd like to thank Yisong Yue for the opportunity, the TAs, and the institute for providing us this opportunity \& their feedback.
\newpage
\ENDDOCUMENT
\newpage
\section*{Appendix}
\textbf{LLM Prompts \& Outputs:}
\begin{verbatim}
You are a music analysis assistant  
that creates optimized prompts for  
SUNO AI based on the user's  
music taste. The user will  
provide one or more song  
links or titles. Analyze the  
provided songs deeply — considering their  
beats (tempo, rhythm, syncopation), instrumentation  
(types of instruments used and  
how they are layered), tone  
and mood (emotional feel, e.g.  
happy, dark, melancholic, energetic, relaxed),  
style (genre/sub-genre, cultural influences,  
vocal style), production qualities (analog/digital,  
lo-fi, clean, lush, minimalistic), melodic  
structure (hooks, vocal phrasing, chord  
progressions), lyrical themes (stories or  
emotions conveyed), and vocal processing  
(natural, auto-tuned, layered, reverb-heavy, etc.).  

Then generate the following:  

(1) a Lyrics Prompt of  
no more than 200 characters  
to guide SUNO AI's lyric  
generation;  

(2) a Style Description with  
1000 char limit, describing in  
detail the musical style and  
feel SUNO AI should aim  
for, mentioning relevant elements such  
as instruments, mood, beat structure,  
production style, genre influences, vocal  
style, and — if the  
LLM deems it important —  
whether a male or female  
voice should be used. Do  
not include any artist names  
in this section.  

(3) Generate a full Song  
Title to be used for  
the song — you may  
include the artist name in  
the Song Title if you  
think it enhances the listener's  
perception of the song or  
aligns with the style, but  
provide only the Song Title  
here with no explanation.  

(4) Provide a Reasoning section  
that explains, in simple and  
clear language, how the provided  
songs influenced your choices. Describe  
what aspects of the songs  
(beats, mood, instrumentation, lyrical themes,  
etc.) appeared most commonly or  
most strongly, what they reveal  
about the user's musical preferences,  
and why you selected this  
particular Lyrics Prompt, Style Description,  
and Song Title based on  
that understanding of the user's  
taste.
\end{verbatim}
The output:
\begin{verbatim}
    (1) Lyrics Prompt:  
A heartfelt journey of love  
and longing, blending hope and  
melancholy with vivid imagery and  
emotional depth.  

(2) Style Description:  
Aim for a genre-blending style  
that fuses modern indie folk  
with ethereal pop elements. The  
instrumentation should feature acoustic guitar  
as a warm, organic base,  
layered with gentle piano and  
subtle string arrangements to create  
lush, cinematic textures. A mid-tempo  
beat with a soft but  
driving rhythm provides steady momentum  
without overpowering the introspective mood.  
The tone is wistful and  
reflective, evoking feelings of nostalgic  
longing and delicate beauty. Vocals  
should be female, intimate and  
emotive, with natural phrasing infused  
with light reverb for spaciousness  
and emotional impact. Production is  
clean with a balance of  
organic acoustic sounds and polished  
digital clarity, emphasizing transparency and  
emotional resonance. Melodic structure centers  
on memorable hooks and simple,  
heartfelt chord progressions supporting expressive  
lyrical delivery.  

(3) Song Title:  
"Leliță Mărie: Echoes of a  
Tender Heart"  

(4) Reasoning:  
The playlist features a diverse  
set of songs spanning indie  
folk ("Little Talks," "Leliță Mărie"),  
electronic remixes ("Shake It To  
The Max (FLY) - Remix,"  
"Faded"), and emotive pop ("I  
Love It," "Fast Car"). The  
inclusion of traditional or folk-tinged  
tracks like "Leliță Mărie" and  
"Frunză Verde Foi Mărunte" reveals  
a preference for music with  
emotional depth and cultural nuance.  
The mix of acoustic instruments  
with modern production styles points  
toward a taste for music  
that blends organic and digital  
elements, with a strong emphasis  
on melody and mood. Lyrics  
tend toward evocative storytelling and  
emotional themes rather than abstract  
or purely dance-focused content. The  
choice for a female vocal  
with natural yet spacious delivery  
reflects this emotional openness. Balanced  
tempo with both reflective and  
rhythmic elements suits the varied,  
yet cohesive feel of the  
user’s favorites. The generated prompt  
and style aim to capture  
that heartfelt, sophisticated vibe combining  
folk warmth and pop accessibility,  
inspired particularly by the standout  
vocal-driven tracks in the list,  
such as "Leliță Mărie" and  
"Little Talks."
\end{verbatim}
\textbf{Source Code: }\href{https://github.com/Jafarbek/CS159.git}{Github}\\
\textbf{LaTeX Source: }Adapted from \href{https://math.arizona.edu/~kirti/}{Kirti Joshi}'s Math preprints.

\end{document}